\documentstyle[12pt]{article}
\def\be{\begin{equation}}
\def\ee{\end{equation}}
\def\bea{\begin{eqnarray}}
\def\eea{\end{eqnarray}}
\def\l{\label}
\def\c{\cite}

\def\ta{\tilde \alpha}
\def\bp{\beta_+}
\def\bm{\beta_-}
\def\p{\partial}

\begin{document}
\begin{titlepage}


\vspace{2cm}

\begin{center}
\Large
{\bf  Stringy Quantum Cosmology of the Bianchi Class A}

\vspace{2cm}

\normalsize

\large{James E. Lidsey}

\normalsize
\vspace{1cm}

{\em NASA/Fermilab Astrophysics Center, \\
Fermi National Accelerator Laboratory, Batavia, IL 60510}

\vspace{3.5cm}

\end{center}

\baselineskip=24pt
\begin{abstract}
\noindent

The quantum cosmology of the string effective action is considered
within the context of the Bianchi class A minisuperspace. An exact
unified solution is found for all Bianchi types and interpreted
physically as a quantum wormhole. The solution is generalized for
types ${\rm VI_0}$ and ${\rm VII_0}$. The Bianchi type IX wavefunction
becomes increasingly localized around the isotropic Universe at large
three-geometries.

\end{abstract}

\vspace{1cm}


\vspace{1cm}

\small{$^1$Electronic address - jim@fnas09.fnal.gov}

\normalsize

\end{titlepage}

\setcounter{equation}{0}


The ongoing experiments investigating the cosmic microwave background
radiation indicate that the Universe is very nearly isotropic on very
large scales and the origin of this observed isotropy remains a
fundamental problem in modern cosmology \c{COBE}. The inflationary
paradigm offers a partial resolution in the form of the cosmic no hair
conjecture \c{GH1977}; a Universe dominated by vacuum energy should
rapidly approach the de Sitter solution regardless of initial
conditions and any primordial anisotropies and inhomogeneities are
therefore washed out. Such an explanation is not complete, however,
since there exist solutions that recollapse before the vacuum energy
is able to dominate. This suggests that additional physics is required
and further insight might be gained at the Planck epoch where quantum
gravity effects are thought to be important.

The superstring theory \c{string} and the quantum cosmology program
\c{WDW} are two approaches to the subject of quantum gravity that have
been investigated in some detail. If the superstring is indeed the
ultimate `theory of everything', it must explain the observed isotropy
at some level. Unfortunately a consistent quantum field theory of the
superstring is not currently available and general solutions to the
full Wheeler-DeWitt (WDW) equation have yet to be found. However,
progress can be made by making a minisuperspace approximation and
considering the quantum cosmology of the string effective action
\c{quantumstring}.  Recently an exact solution for the Bianchi IX
minisuperspace was found by Lidsey and remarkably the wavefunction
becomes peaked around the isotropic solution as the three-volume of
the Universe increases \c{lidsey}. This represents the first,
non-vacuum, exact Bianchi IX solution to the WDW equation and the
purpose of this essay is to derive this solution and generalize it
within the context of the Bianchi class A models.

The spatially homogeneous models admit a Lie group $G_3$ of isometries
transitive on spacelike three-dimensional orbits. The Euclidean
$4$-metric is
\be
\l{4metric} ds^2=dt^2+ h_{ab} \omega^a \omega^b,
\qquad a,b=1,2,3,
\ee
where the 3-metric on the surfaces
of homogeneity is given by
\be
h_{ab}(t)=e^{2\alpha (t)} \left(
e^{2\beta (t)} \right)_{ab}
\ee
and the matrix $\beta_{ab} =
{\rm diag} \left[ \bp +\sqrt{3} \bm , \bp -\sqrt{3}\bm , -2\bp
\right]$ is traceless.  The one-forms $\omega^a$ satisfy the
Maurer-Cartan equation $d\omega^a =\frac{1}{2}{C^a}_{bc}\omega^b
\wedge \omega^c$, where ${C^a}_{bc} = m^{ad} \epsilon_{dbc} +
{\delta^a}_{\left[ b \right.} a_{\left. c\right] }$ are the structure
constants of the Lie algebra of $G_3$, $a_c \equiv {C^a}_{ac}$ and
$m_{ab}$ is symmetric. The Jacobi identity ${C^a}_{b\left[ c \right.}
{C^b}_{\left. de \right]} =0$ implies that $a_b$ is transverse to
$m^{ab}$, i.e.  $m^{ab}a_b=0$, and the Lie algebra belongs to the
Bianchi class A if $a_b=0$ \c{A}. This class consists of types I, II,
${\rm VI_0}$, ${\rm VII_0}$, VIII and IX and the Lie algebra of each
type is uniquely determined up to isomorphisms by the rank and
signature of $m^{ab}$.

There are no closed topologies in the Bianchi class B and this may
explain why a standard Hamiltonian treatment is not available for this
class \c{B}.  Consequently we restrict our attention to the class A.
The spatially flat $(k=0)$ and spatially closed $(k=+1)$ Friedmann
Universes are the isotropic limits $(\bp =\bm = 0)$ of the Bianchi
types $\{{\rm I}, {\rm VII_0}\}$ and IX respectively.

We take as our starting theory the bosonic sector of the effective
Euclidean action of the heterotic string in critical (ten) dimensions
to zero-order in the inverse string tension \c{string}. After
compactification onto a six-torus the dimensionally reduced,
four-dimensional action is equivalent to a scalar-tensor theory if the
field strength of the antisymmetric tensor vanishes. In this case the
simplest form of the WDW equation is derived by performing a conformal
transformation $\tilde{g}_{\mu\nu}=\Omega^2g_{\mu\nu}$ on the 4-metric
in such a way that the action is rewritten as Einstein gravity
minimally coupled to a set of massless scalar fields. If
$\Omega^2=e^{-\phi}$, where $\phi$ is the shifted dilaton, the
transformed Euclidean action becomes
\be
\l{actionconf}
S=\int d^4 x
\sqrt{\tilde{g}} \left\{ - \tilde{R} +\frac{1}{2} \left(
\tilde{\nabla}\phi \right)^2 + \frac{1}{2} \sum_{j=1}^3 \left[ \left(
\tilde{\nabla}\psi_j \right)^2 + e^{-2\psi_j} \left( \tilde{\nabla}
\sigma_j \right)^2 \right] \right\},
\ee
where $\tilde{g} \equiv {\rm
det}\tilde{g}_{\mu\nu}$, $\{\psi_j, \sigma_j\}$ are scalar fields
arising from the compactification and we choose units such that $\hbar
= c= 16 \pi m_p^{-2} \equiv 1$ in the conformal frame \c{KM1993}.  If
we further assume that the dilaton is constant on the surfaces of
homogeneity, the transformed world-interval is $d\tilde{s}^2=d\eta^2+
\tilde{h}_{ab}\omega^a\omega^b$, where \be \l{conf3} \tilde{h}_{ab} =
e^{2\ta} \left( e^{2\beta} \right)_{ab} = e^{-\phi+2\alpha} \left(
e^{2\beta} \right)_{ab} \ee is the rescaled 3-metric, $\eta \equiv
\int dt \Omega (t)$ and $\ta \equiv \alpha + \ln \Omega$.  Since $G_3$
is time-independent, a given Bianchi type is symmetric under the
action of this conformal transformation and we may therefore consider
solutions directly in the conformal frame.

Theory (\ref{actionconf}) has ten degrees of freedom $q^{\mu}=(\ta ,
\beta_{\pm}, \phi, \psi_j, \sigma_j )$ with conjugate momenta
$p_{\mu}=\partial S/\p \dot{q}^{\mu}$. Quantization follows by
identifying these momenta with the operators $p_{\mu} =-i\p /\p
q^{\mu}$ and viewing the classical Hamiltonian constraint as a
time-independent Schr\"odinger equation that annihilates the state
vector $\Psi(q^{\mu})$ for the Universe. It is given by \c{lidsey}
\bea
\l{WD}
\left[ e^{-p \ta}\frac{\p}{\p \ta} e^{p\ta}\frac{\p}{\p
\ta} - \frac{\p^2}{\p \bp^2} - \frac{\p^2}{\p \bm^2} +U\left( \ta ,
\beta_{\pm} \right) -12 \frac{\p^2}{\p \phi^2} \right. \nonumber \\
\left. -12 \sum_{j=1}^3 \left( \frac{\p^2}{\p \psi_j^2} + e^{2\psi_j}
\frac{\p^2}{\p \sigma_j^2} \right) \right] \Psi =0 ,
\eea
where the
superpotential
\be
\l{superpotential}
U=\frac{1}{3} \left[ (n_{33}
\tilde{h}_{33} )^2 +( n_{11}\tilde{h}_{11} - n_{22}\tilde{h}_{22})^2 -
2n_{33} \tilde{h}_{33} ( n_{11} \tilde{h}_{11} +n_{22} \tilde{h}_{22}
) \right]
\ee
is determined by the structure constants of the Lie
algebra and the constant $p$ accounts for ambiguities in the operator
ordering. The non-zero eigenvalues of $n_{ab}$ are summarized in Table
1.

\begin{table}
\begin{center}
\begin{tabular}{c||c|c|c|c}

Type &$n^{11}$ &$n^{22}$ &$n^{33}$ &$|S|$ \\
\hline
\hline
& & & & \\
I &0 &0 &0 &0 \\
& & & & \\
II &0 &0 &1 &$\frac{1}{6} e^{2\ta -4\bp}$ \\
& & & & \\
${\rm VI}_0$ &-1 &1 &0 &$\frac{1}{3}e^{2\ta +2\bp} \sinh
2\sqrt{3} \bm$ \\
& & & & \\
${\rm VII}_0$ &1 &1 &0
&$\frac{1}{3}e^{2\ta +2\bp} \cosh 2\sqrt{3} \bm$ \\
& & & & \\
VIII &1
&1 &-1 &$\frac{1}{6}e^{2\ta} \left[-e^{-4\bp}+2e^{2\bp}\cosh 2\sqrt{3}
\bm \right]$ \\
& & & & \\
IX &1 &1 &1 &$\frac{1}{6}e^{2\ta}
\left[e^{-4\bp}+2e^{2\bp}\cosh 2\sqrt{3} \bm \right]$

\end{tabular}
\end{center}

\footnotesize {\hspace*{0.2in} Table 1: The non-zero eigenvalues of
$n^{ab}$ for the Bianchi class A are shown for each Bianchi type. This
matrix has the same rank and signature for a given Bianchi type as the
matrix determining the structure constants of the Lie algebra of
$G_3$.  The function $S= \pm \frac{1}{6} n^{ab} \tilde{h}_{ab}$, where
summation over upper and lower indices is implied, is a solution of
the Euclidean Hamilton-Jacobi equation (\ref{super}). It is
interpreted physically as the conformally transformed Euclidean action
for the classical vacuum theory. The reader is referred to the text
for details.}

\end{table}

Although solving this equation appears to be a formidable task, one
gains valuable insight from the Hamiltonian ${\cal{H}}_{g}$ of the
vacuum theory. At the classical level the Universe may be thought of
as a zero-energy point particle moving in a time-dependent potential
well with $ 0= \l{classHam} {\cal{H}}_{g} \propto
G^{\lambda\kappa}p_{\lambda}p_{\kappa} +U(q^{\lambda})$, where
$G^{\lambda\kappa}$ is the $(2+1)$-Minkowski space-time metric, $\ta$
plays the role of `time' and $\{\lambda ,\kappa \}$ represent the $\ta
,\beta_{\pm}$ minisuperspace coordinates. This implies that the vacuum
Bianchi A minisuperspace can be supersymmetrized \c{super} by solving
the Euclidean Hamilton-Jacobi equation
\be
\l{super}
U =
G^{\lambda\kappa} \frac{ \p S}{\p q^{\lambda}} \frac{\p S}{\p
q^{\kappa}}
\ee
and introducing the fermionic variables
$\chi^{\kappa}, \bar{\chi}^{\lambda} $ defined by the spinor algebra
\be
\l{algebra}
\chi^{\kappa} {\chi}^{\lambda} + {\chi}^{\lambda}
\chi^{\kappa}= \bar{\chi}^{\kappa} \bar{\chi}^{\lambda} +
\bar{\chi}^{\lambda} \bar{\chi}^{\kappa} =
\bar{\chi}^{\kappa}\chi^{\lambda} + \chi^{\kappa}\bar{\chi}^{\lambda}
- G^{\kappa\lambda} =0 .
\ee
The Hamiltonian is then equivalent to
${\cal{H}}_g \propto Q\bar{Q}+\bar{Q}Q$, where the supercharges
\be
\l{S}
Q=\chi^{\kappa} \left( p_{\kappa} +i \frac{\p S}{\p q^{\kappa}}
\right), \qquad \bar{Q}=\bar{\chi}^{\kappa} \left( p_{\kappa}
-i\frac{\p S}{\p q^{\kappa}} \right)
\ee
satisfy $Q^2= 0= \bar{Q}^2$,
and after quantization one obtains the `square roots' of the WDW
equation, i.e. $Q\Psi = 0= \bar{Q}\Psi$.  It follows that a solution
to the standard WDW equation may be found by solving these square
roots and restricting one's attention to the bosonic sector of the
wavefunction. It is straightforward to show that, modulo a constant of
proportionality, the bosonic component of the wavefunction annihilated
by the supersymmetric quantum constraints is \c{super}
\be
\l{boson}
\Psi_{\rm bosonic} =e^{-S} .
\ee

We find that a unified solution to Eq. (\ref{super}) is
\be
\l{supersolution}
S= \pm \frac{1}{6} n^{ab} \tilde{h}_{ab}
\ee
where
summation over upper and lower indices is implied and the full
expressions for each Bianchi type are presented in Table 1. The
elegance of (\ref{boson}) and (\ref{supersolution}) motivates us to
separate the matter and gravitational sectors of the WDW equation
(\ref{WD}) with the ansatz $\Psi =X(\ta ,\beta_{\pm} )Y (\phi , \psi_j
, \sigma_j )$, where $X=W(\ta ) e^{-S (\tilde{\alpha} , \beta_{\pm})
}$. This implies that
\be
\l{WD1}
\left[ \frac{\p^2}{\p \ta^2}
+p\frac{\p}{\p \ta} - \frac{\p^2}{\p \bp^2} - \frac{\p^2}{\p \bm^2} +U
-z^2 \right]X=0
\ee
and
\be
\l{WD2} \left[ \frac{\p^2}{\p \phi^2}
+\sum_{j=1}^3 \left( \frac{\p^2}{\p \psi_j^2} +e^{2\psi_j}
\frac{\p^2}{\p \sigma_j^2} \right) -\frac{z^2}{12} \right] Y=0,
\ee
where $z$ is an arbitrary separation constant that can be interpreted
physically as the total momentum eigenvalue of the matter sector. This
separation is possible for any theory that is equivalent at the
classical level to Einstein gravity minimally coupled to a stiff
perfect fluid.

It follows from Eq. (\ref{supersolution}) that $\p S/\p \ta =2S$ and
$G^{\kappa \lambda} \p^2 S/\p q^{\kappa} \p q^{\lambda} =12 S$ for all
Bianchi types. We therefore deduce with the help of these identities
that
\be
\l{X}
X=e^{(3 - p/2) \ta -S}
\ee
is a solution to Eq.
(\ref{WD1}) provided we choose $p^2=4(9-z^2)$. It only remains to
solve Eq.  (\ref{WD2}) and this is achieved with the separable ansatz
\be
Y= A_1 (\psi_1) A_2(\psi_2) A_3 (\psi_3) e^{\pm i\gamma\phi \pm
i(\omega_1\sigma_1 + \omega_2\sigma_2 + \omega_3\sigma_3 )}
\ee
where
$A_j$ satisfy the wave equation of Liouville quantum mechanics and
$\{\gamma ,\omega_j \}$ are separation constants \c{lidsey}. A
reduction to the Bessel equation follows after the change of variables
$\psi_j =\ln \xi_j$ and this yields the general solutions $A_j =
Z_{\pm \Lambda \cos \theta_j}(\omega_j e^{\psi_j})$, where $Z$ denotes
some linear combination of modified Bessel functions of the first and
second kinds, $\Lambda^2=\gamma^2 +z^2/12$ and the constants
$\theta_j$ are solutions to the constraint equation $\sum_{j=1}^3
\cos^2 \theta_j =1$.

Finally the solution in the original frame follows from Eq.
(\ref{conf3}). We therefore arrive at the unified set of solutions
\bea
\l{WDsolution}
\Psi =\exp \left[ \left( 3-\frac{p}{2} \right)
\alpha -\frac{1}{6} e^{-\phi} \left| n^{ab} h_{ab} \right| + \left(
\frac{1}{4} \left( p - 6 \right) \pm i \gamma \right) \phi \right]
\nonumber \\
\times \prod_{j=1}^3 \left[ Z_{\pm \Lambda \cos \theta_j}
\left( \omega_j e^{\psi_j} \right) e^{\pm i \omega_j \sigma_j}\right].
\eea
These solutions cannot be interpreted as Lorentzian
four-geometries because the wavefunction is Euclidean for all values
of the scale factor. On the other hand, they remain regular, in the
sense that they do not oscillate an infinite number of times, when the
spatial metric degenerates $(\alpha =-\infty)$ and, with the exception
of the type I solution, they are exponentially damped at large
$\alpha$. Consequently they satisfy the Hawking-Page boundary
conditions and may therefore be interpreted as quantum wormhole
solutions \c{HP1990}.

The function $W$ behaves as a variable amplitude for the gravitational
component of the wavefunction. The type ${\rm VI}_0$ and ${\rm VII}_0$
solutions may be generalized in such a way that this amplitude becomes
a function of both the $\tilde{\alpha}$ and $\beta_+$ variables.  We
find that a second solution to Eq. (\ref{WD1}) is
\be
X= \exp \left[
\frac{1}{24} \left( ( p^2 -12p +4z^2 +36 ) \alpha + ( p^2 -36 +4z^2 )
\beta_+ \right) -S \right]
\ee
and these solutions do not depend on a
specific choice of factor ordering.

The problem of extracting physical predictions in quantum cosmology
from the wavefunction of the Universe is an unresolved one. However,
it is reasonable to suppose that a strong peak in the wavefunction
represents a prediction in some sense. This is the case, for example,
if one adopts the Hartle-Hawking proposal and interprets $|\Psi|^2$ as
an unnormalized probability density \c{HH1983}. When the three-surface
degenerates, $S$ becomes vanishingly small for all Bianchi types, and
the wavefunction exhibits no peak in the $(\bp ,\bm )$ plane. As shown
in Figure 1, however, the Bianchi IX wavefunction becomes strongly
localized around the isotropic solution $\bp = \bm =0$ as the scale
factor increases. This suggests that there is a progressively higher
probability of finding this Universe in the isotropic state as
$\alpha$ increases \c{MR1991}.  Unfortunately the same is not true for
the Bianchi type ${\rm VII}_0$. We see from Figure 2 that the
wavefunction is indeed peaked around $\beta_-=0$, but there is no
local maximum for finite $\beta_+$. Consequently, it is not clear
whether this solution selects the spatially flat Friedmann Universe.

In conclusion we have found a unified exact solution to the WDW
equation for the Bianchi class A minisuperspace with a matter sector
motivated by the string effective action. If such a solution is to
have any direct cosmological relevance, it is necessary to find a
mechanism that leads to the emergence of the classical domain. In
principle, such a domain could be reached if one or more of the scalar
fields were to acquire an effective mass once the scale factor became
sufficiently large. In the type IX example the Universe would then
tunnel from the Euclidean regime into a highly isotropic Lorentzian
state and this may point towards a possible non-inflationary
resolution of the isotropy problem.

\vspace{0.5cm} {\bf Acknowledgments} The author is supported by a
Science and Engineering Research Council (SERC) UK postdoctoral
research fellowship and is supported at Fermilab by the DOE and NASA
under Grant NAGW-2381.

\newpage

\phantom{figure}

\vspace{3.1in}

{\em Figure 1:} The gravitational component of the Bianchi type IX
wavefunction (\ref{WDsolution}) is plotted when $\ta = 0$. The
wavefunction is localized around the point $\beta_+=\bm =0$ in the
$(\bp ,\bm )$ plane. This point corresponds to the isotropic Friedmann
solution. The peaked becomes more pronounced as the scale factor
increases.

\vspace{3.5in}

{\em Figure 2:} The equivalent solution to Figure 1 for the Bianchi
type ${\rm VII}_0$ cosmology. The wavefunction is peaked around $\bm
=0$, but there is no local maximum for finite $\beta_+$.

\end{document}